\documentclass[runningheads]{svmult}
\usepackage{subfigure}
\usepackage{makeidx}   
\usepackage{graphicx}  
\usepackage{subeqnar}  
\usepackage{multicol}  
\usepackage{physprbb}  
\makeindex             

\begin{document}

\title*{Universality Classes for Force Networks in Jammed Matter}

\author{Srdjan Ostojic \and Bernard Nienhuis}
 \institute{Institute for
Theoretical Physics, Universiteit van Amsterdam, Valckenierstraat 65,
1018 XE Amsterdam, The Netherlands}

\maketitle

\begin{abstract}
We  study the geometry  of forces in  some simple  models for
granular   stackings.  The  information   contained  in   geometry  is
complementary  to that  in  the  distribution of  forces  in a  single
inter-particle  contact, which  is more  widely studied.  We  present a
method which focuses  on the fractal nature of  the force network and
find  good evidence  of scale  invariance.  The method  enables us  to
distinguish    universality   classes   characterized    by   critical
exponents. Our approach can  be applied to force networks in other
athermal jammed systems.
\end{abstract}

\section{Jammed Matter and Force Networks}

Aggregates of particles can be  found in a disordered solid-like state
resulting   from    the   phenomenon   of    jamming   \cite{jam:book,jam:liu-nagel0,jam:weitz,jam:makse1,jam:bouchaud}.    Granular
materials, colloidal  suspensions and molecular liquids are  but a few
examples of  such systems that  present a non-zero yield  stress while
trapped  in one  of  many accessible  metastable  states.  If  thermal
fluctuations  are  irrelevant,  the   forces  on  each  particle  must
balance. Each  stable configuration is thus characterized  by a highly
irregular network of forces spanning the entire system.

Experimental  \cite{fluct:mueth,fluct:nor,fluct:blair1,fluct:erikson,fluct:brujic2}                      and                      numerical
\cite{fluct:radjai1,fluct:luding,fluct:radjai2,fluct:makse,force:copper2,fluct:antony,jam:liu-nagel,jam:liu-nagel2,fluct:silbert}  studies  have  identified
two  main  distinctive features  of  these  force networks.   Firstly,
strong  fluctuations are  found  in the  magnitudes of  inter-particle
forces.   The  associated distribution  function  $P(F)$ displays  two
characteristic properties: (i) it decays exponentially at large forces
and (ii)  it exhibits a plateau  or small peak at  small forces, which
has  been   identified  as  a   signature  of  jamming.    The  second
experimental observation  is that large forces  are concentrated along
tenuous paths,  which have been deemed ``force  chains''. While $P(F)$
has  been commonly  used also  as  a characterization  of these  force
chains, strictly speaking it provides no information about the spatial
organization  of forces.   In  fact,  so far  force  chains have  been
identified mainly visually,  and a quantitative characterization seems
to be lacking.

By drawing  an analogy with percolation,  in this Letter  we develop a
geometrical description  which associates a set  of critical exponents
with an ensemble  of force networks.  We apply  this approach to three
different models of static  granular media under uniform pressure.  We
find that  they belong  to different geometrical  universality classes
although $P(F)$ displays similar features in all three of them.

\section{Force  clusters}

  Consider an ensemble  of  configurations
of   a   fixed   number   of jammed particles,   obtained   numerically   or
experimentally.  Each configuration defines a contact graph $G$, where
nodes correspond  to particle centers  and edges connect  particles in
contact.  Assuming there is no friction, the inter-particle forces are
normal to the  particle surface, and the underlying  force network can
be  represented  by  associating  with   each  edge  $i$  of  $G$  the
corresponding force  magnitude $F_i$.  To investigate  the geometry of
forces, rather then  the underlying geometry of contacts,  we choose a
threshold $f$ and  look at the subgraph ${\bar  G(f)}$ of $G$ obtained
by selecting only the edges with $F_i>f$. For $f$ small, ${\bar G(f)}$
consists of a single connected component, but as $f$ increases, ${\bar
G(f)}$ breaks up  into a number of disconnected  clusters. An ensemble
of force  networks thus induces a family  of probability distributions
of cluster sizes $\rho(s,f)$ for different thresholds $f$, the cluster
size $s$ being defined as the number of edges in a cluster.

If  the forces  $F_i$  were distributed  independently  for each  $i$,
e.g.  uniformly between  $0$ and  $1$, then  the force  clusters would
simply be  bond percolation clusters \cite{stauff}.  In  that case, in
the thermodynamic limit $N\to \infty$,  a phase transition occurs at a
critical  value  $f_c$  of   $f$:  an  infinite  cluster  exists  with
probability $1$ for $f<f_c$, and  with probability $0$ for $f>f_c$. At
$f_c$,    the    cluster     sizes    are    power-law    distributed,
$\rho(s,f_c)\propto s^{-\tau}$, and the correlation length diverges as
$\xi  \propto  |f-f_c|^{-\nu}$   near  the  threshold.   The  scaling
exponents $\tau$ and $\nu$ are  universal, they are independent of the
underlying  geometry, and  in fact  they do  not depend  on  the local
distribution of forces  $P(F)$ or even their correlations,  as long as
these are short-ranged.

In an  ensemble of  force networks corresponding  to a  jammed system,
force  and  torque  balance  on  each particle  cause  dependence  and
long-range correlations  between bonds.  Nevertheless,  if the average
forces are uniform over the extent  of the system, we expect to find a
critical threshold  $f_c$ and an associated set  of universal scaling
exponents.  The analogy with  percolation moreover suggests that these
exponents  are   independent  of  $P(F)$  and  thus   provide  a  new,
complementary characterization of force networks.

 \begin{figure}[h]
\begin{center}
\begin{minipage}{0.4\linewidth}
\includegraphics[width=0.8\linewidth]{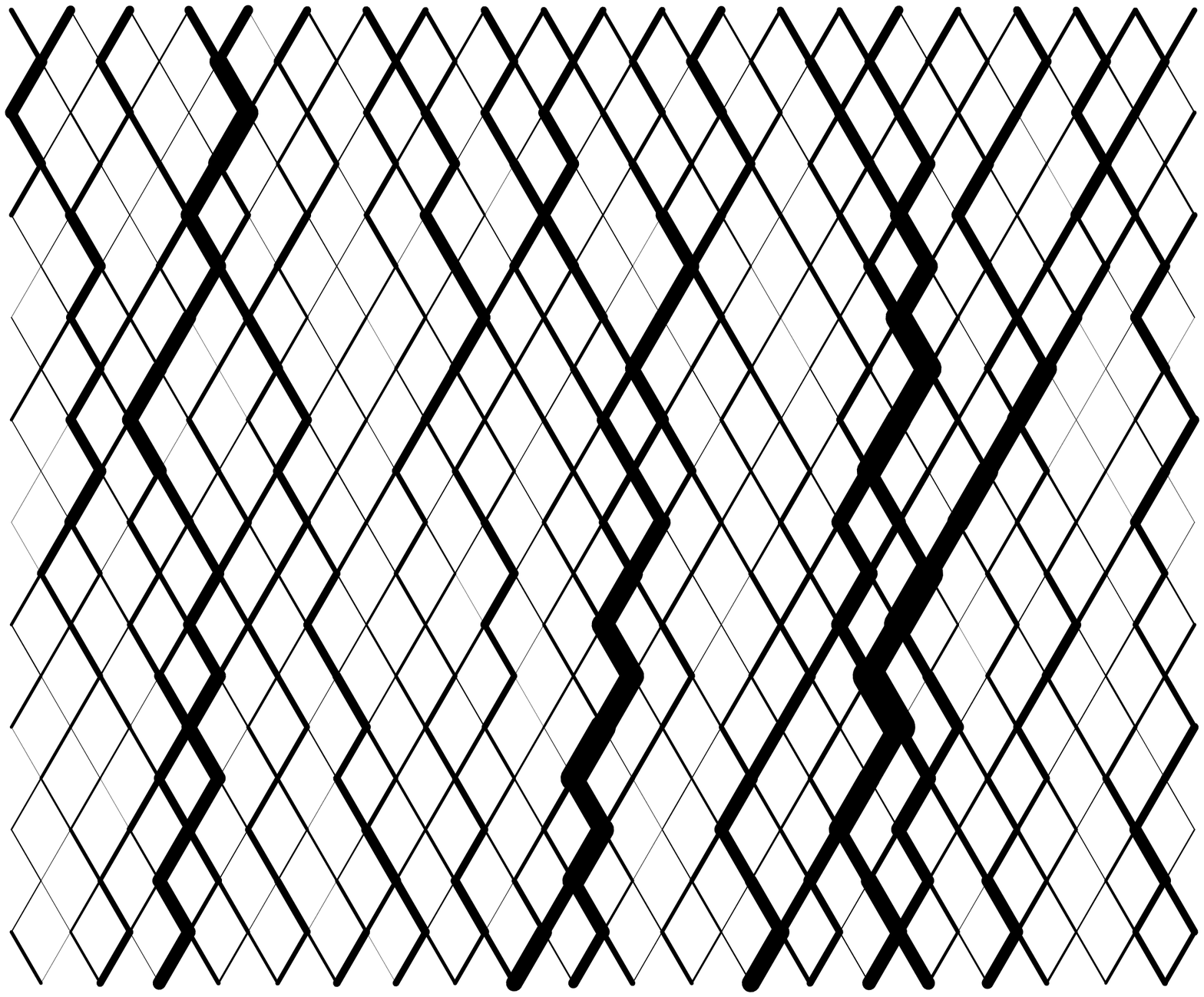}
\end{minipage}
\begin{minipage}{0.4\linewidth}
\includegraphics[width=0.8\linewidth]{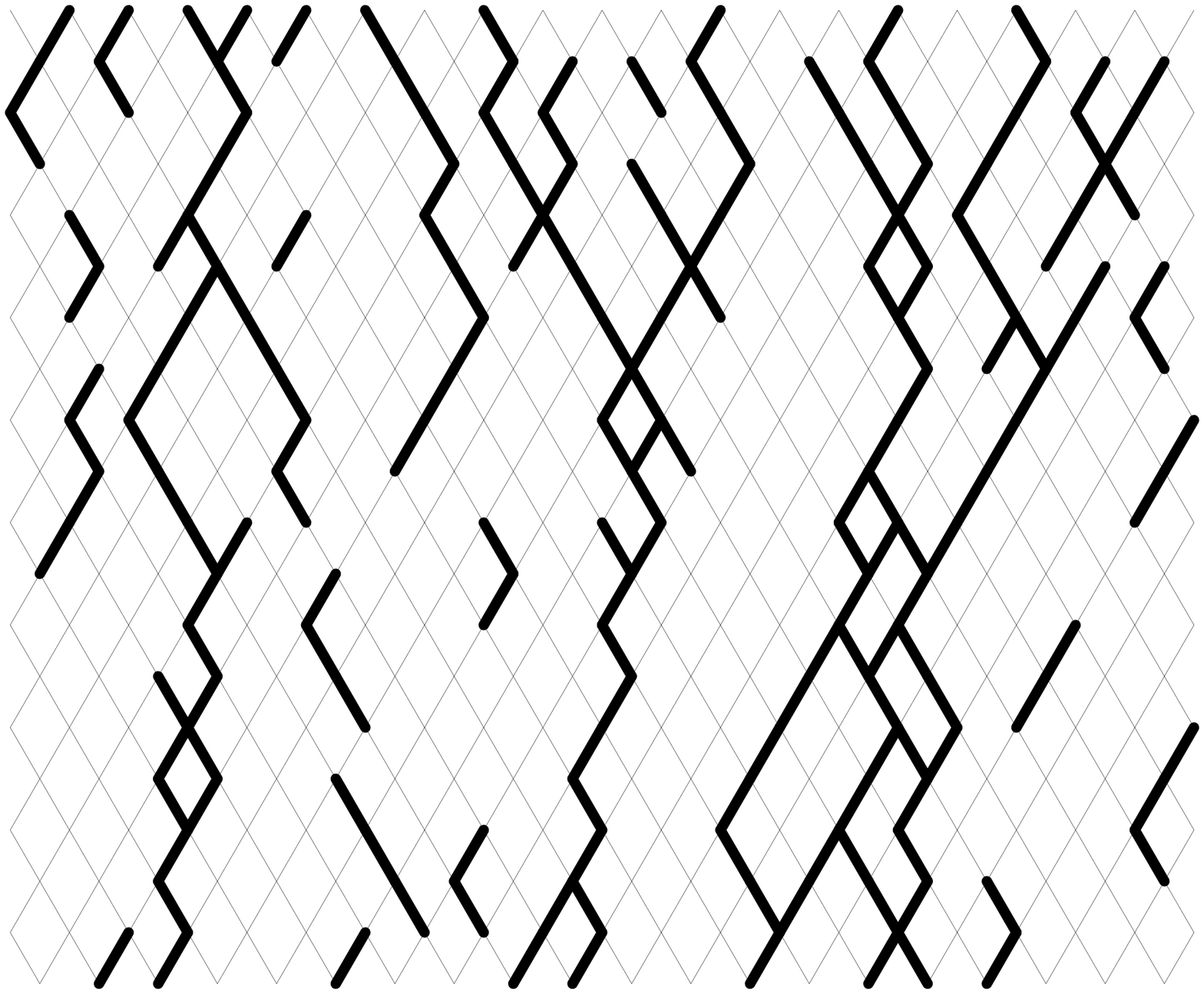}
\end{minipage}
\begin{minipage}{0.4\linewidth}
\includegraphics[width=0.8\linewidth]{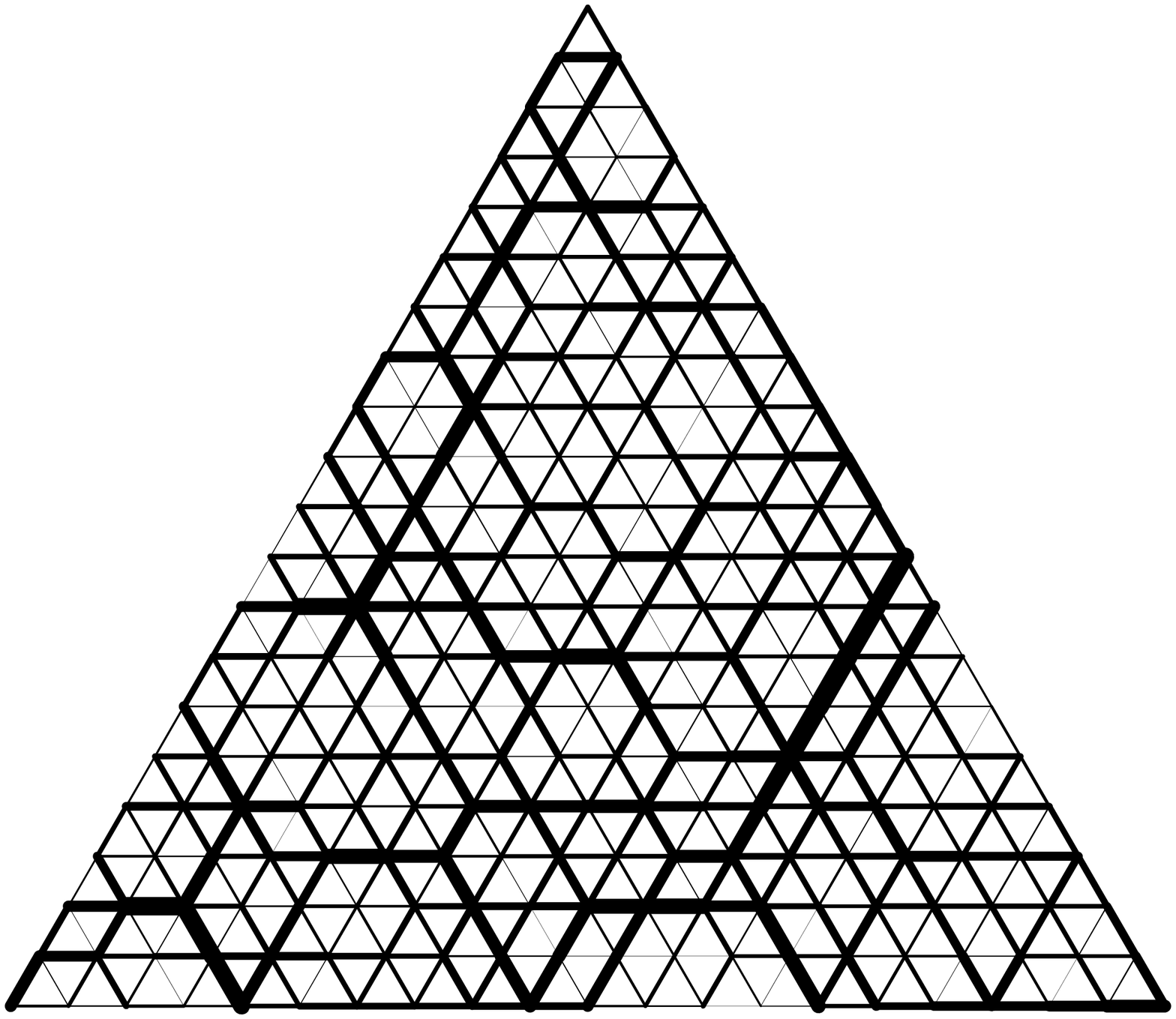}
\end{minipage}
\begin{minipage}{0.4\linewidth}
\includegraphics[width=0.8\linewidth]{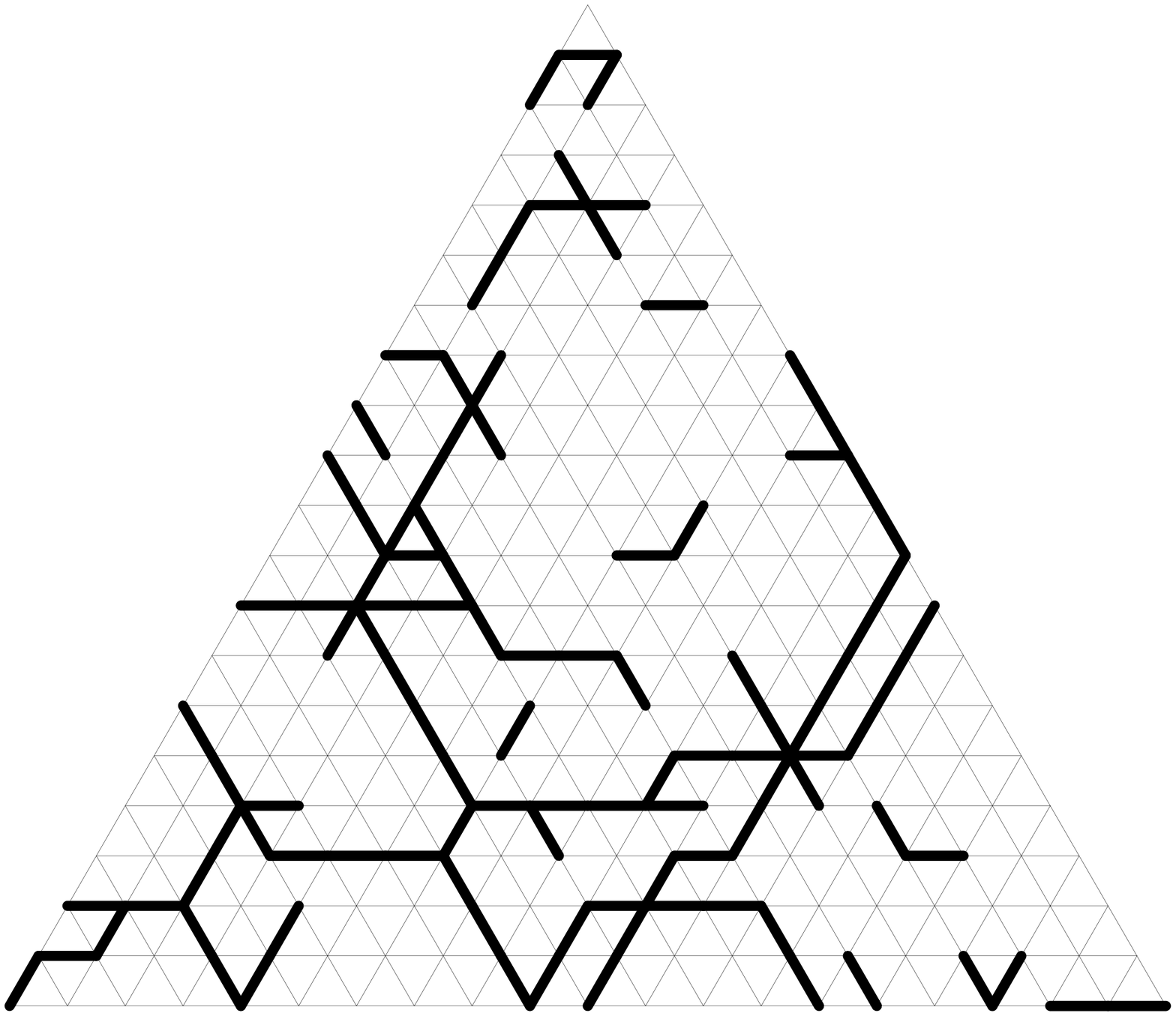}
\end{minipage}
\end{center}
\caption{Examples of force networks (the thickness of the lines is
  proportional to the force magnitude) and corresponding force clusters
  close to the critical threshold:
 packing of  400 grains in Model
  A (top) and  packing of 200 grains in Model B (bottom). }\label{models}
\end{figure}  

\section{Criticality and finite size  scaling.} 
An  efficient method  is necessary  to  study the  existence of  scale
 invariance   around  the  critical   threshold  from   numerical  and
 experimental data. While, strictly speaking, the system becomes scale-invariant
 only in  the thermodynamic limit,  $f_c$ and the  associated critical
 exponents can be extracted from  data on systems of finite size using
 finite  size scaling \cite{fss}.   This describes  the scaling  of an
 observable with the  system size close to criticality:  if a quantity
 $X$  is expected  to diverge  as $|f-f_c|^{-\chi}$  near $f_c$  in an
 infinite system, then  in a system of size $N$,  it obeys the scaling
 law
\begin{equation}
X(N,f)= N^{\phi}{\tilde X}((f-f_c){N}^{1/d\nu})
\label{ffs}
\end{equation} 
with $d$  the spatial  dimension and $\phi=\chi  / d\nu$.  The scaling
function  ${\tilde   X}$  depends   on  a  single   rescaled  variable
$x=(f-f_c){N}^{1/d\nu}$, and for $x\gg  1$ it behaves as $x^{-\chi}$, while
for $x\to 0$ it remains finite.

Using measurements of  $X$ in systems of finite  sizes, the parameters
$\phi$,  $\nu$ and  $f_c$  can  be obtained  from  (\ref{ffs}) in  two
steps. Assuming that $X({N},f)$ as function of  $f$ displays a maximum
$X_m({N})$,  from (\ref{ffs}) the  maxima for different  ${N}$ all
correspond to the same maximum of ${\tilde X}$, hence $X_m({N})\propto
{N}^{\phi}$. Plotting  the amplitudes of  the maxima versus  ${N}$, we
get the  exponent $\phi$. The  values of $f_c$  and $\nu$ can  then be
obtained by  determining the best  data collapse in the  region around
the maximum.

\section{Models studied}

Combining the finite-size scaling method with Monte-Carlo simulations,
we studied  force-cluster criticality in  three two-dimensional models
of     static    granular     matter     under    uniform     pressure
\cite{jaeger:rev,bouch:rev}.  As  all three  models -- which  we will
call  A, B,  and  C for  further  reference --  have been  introduced
earlier in other  contexts, here we only define  them briefly, without
motivating in detail  their relevance to granular matter.  In our view,
they  are   are  the  simplest  implementations   of  two  fundamental
ingredients of force networks, namely  force balance on each grain and
force randomness.

\subsection{Snooker model}
To  start  with, we  consider  the  ``snooker-triangle  packing''  studied  in
\cite{fluct:jacco1,fluct:jacco2}.  It consists of a hexagonal packing
of frictionless spherical grains  confined within a triangular domain,
with the same confining pressure applied on all sides of the triangle.
A  force network  on  this  packing consists  of  repulsive forces  in
vectorial  balance  on each  grain  and  consistent  with the  applied
pressure.   These   constraints  however   do  not  define   a  single
configuration of forces, but a  whole  set. Following  Edwards'
prescription \cite{edwards1}, all such  force networks are taken to be
equally likely,  similarly to  a micro-canonical ensemble.   We sample
this ensemble  with a Metropolis algorithm,  using the parametrization
of force networks developed  in Ref.  \cite{pg}.  In Fig. \ref{models}
we  show  an  example  of  a  force network  in  this  model  and  the
corresponding force clusters for a threshold $f=0.94$.

\subsection{Independent $q$-model \label{modelA}}

   We  next  consider consider  the  scalar  $q$-model
\cite{qmodel}, one of  the first models introduced to  account for the
fluctuations of  forces and appearance  of force chains in  a granular
packing.  Here we consider the massless $q$-model on a periodic tilted
square lattice, which  can be interpreted as a  packing of rectangular
bricks \cite{raj}.   A uniform pressure is  applied on the  top of the
packing and  on each  site a brick  supports a weight  $W_{ij}$. Each
brick transfers {\it vertical} forces $F_l^{(ij)}$ and $F_r^{(ij)}$ to
its  bottom left  and  right neighbors  respectively.  Vertical  force
balance  is automatically  satisfied by  considering  $F_l^{(ij)}$ and
$F_r^{(ij)}$  respectively  as fractions  $q_{ij}$  and $1-q_{ij}$  of
$W_{ij}$, and  randomness in force transfer is  implemented by taking
the $q_{ij}$ uniformly distributed  between $0$ and $1$, independently
for each  site.  Fig.  \ref{models}  shows a force network  in this
model and the corresponding force clusters for a threshold $f=0.7$
(for unit external pressure).

\subsection{Microcanonic $q$-model}
Our third model is a variation on the $q$-model.  We consider the same
packing  as   in  Sec.  \ref{modelA},  but   now,  following  Edwards'
prescription,  all allowed  force networks  -- consisting  of  sets of
vertical forces $\{(F_l^{(ij)},  F_r^{(ij)})\}$ -- are equally likely.
As  shown in  Ref.  \cite{jstat},  this  is equivalent  to having  the
$q_{ij}$   distributed  with   the   joint  probability   distribution
$\prod_{ij} W_{i,j}$.  The aim is to examine the influence of the form
of   the  probability  distribution   by  comparing   independent  and
microcanonic  $q$-models,  and   the  difference  between  scalar  and
vectorial conservation laws by comparison with the snooker model.

\section{Results}
\begin{figure}
\begin{center}
\includegraphics[width=0.8\linewidth]{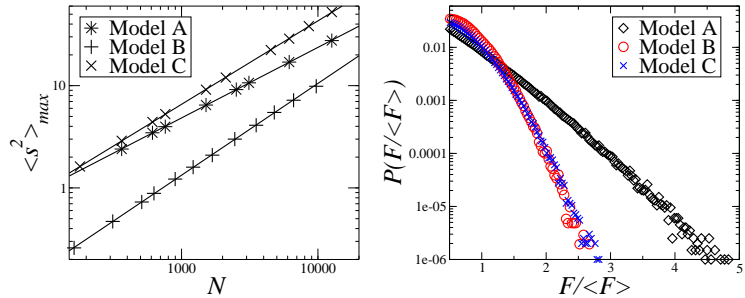}
\end{center}
\caption{Results  of  Monte-Carlo  simulations  for the  three  models
defined in  the text: (a) scaling  of the maxima of  the second moment
$\langle s^2({N},  f) \rangle$ of the distribution  of cluster sizes
(omitting the largest cluster  in every configuration), as function of
the  total system size  $N$; (b)  probability distributions  $P(F)$ of
force  magnitudes, obtained from  $100$ samples  of systems  of $10^4$
particles.}\label{phi}
\end{figure}

A  convenient  observable  to  study  is  the  second  moment  of  the
distribution of  cluster sizes, $\langle s^2({N},  f) \rangle=\int s^2
\rho(s,f_c)$, where the contribution  from the largest cluster in each
configuration is  omitted, and the system  size $N$ is  defined as the
total number  of edges.   In all three  models defined above,  we find
that $\langle s^2({N},  f) \rangle$ displays a maximum  as function of
$f$ for  fixed $N$. The amplitudes  of the maxima as  functions of $N$
follow sharp power-laws shown  in Fig.  \ref{phi} (a), thus confirming
the   existence  of  a   critical  threshold   in  each   model.   The
corresponding   critical   exponent   $\phi$   is  related   via   the
hyper-scaling relation  \cite{stauff} to  $\tau$, the exponent  of the
cluster-size  distribution  at   criticality,  and  $D$,  the  fractal
dimension        of       the        incipient        cluster       as
$\phi=\frac{3-\tau}{\tau-1}=D-1$. Higher  moments $\langle s^n({N}, f)
\rangle    $    display    a    similar   scaling    with    exponents
$\phi_n=\frac{n+1-\tau}{\tau-1}$, implying  that the full distribution
$\rho(s,f_c)$ approaches a scaling form around the critical threshold.

\begin{figure*}
\begin{center}
\includegraphics[width=0.98\linewidth]{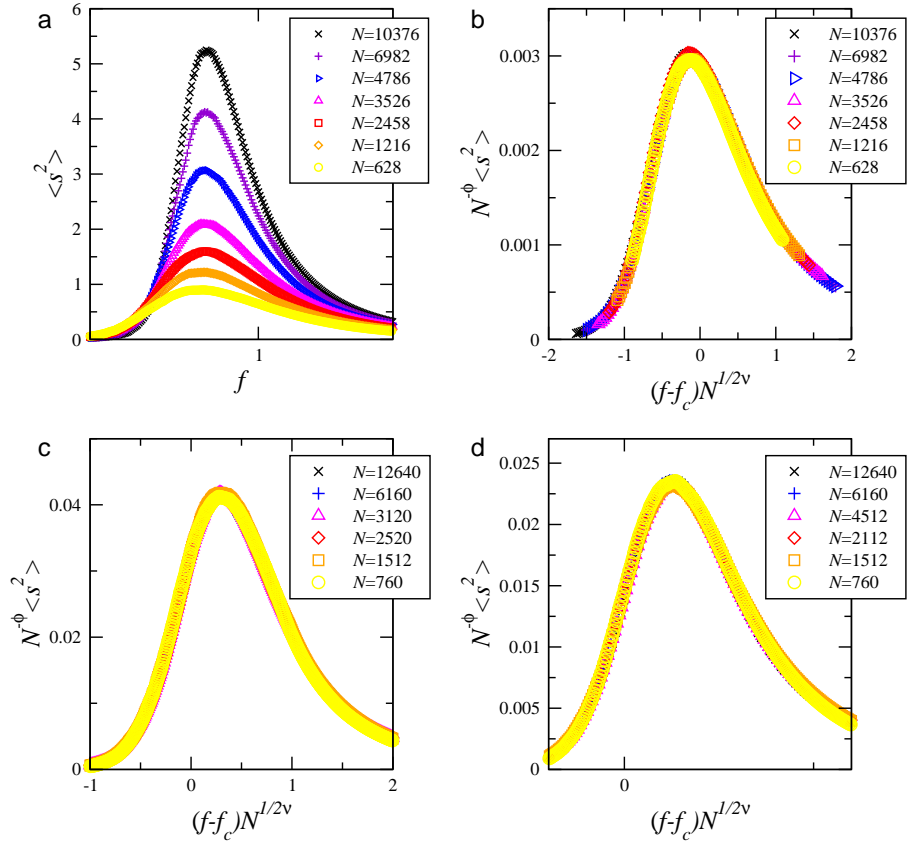}
\end{center}
\caption{ Scaling  of the second moment (omitting  the largest cluster
in each configuration) as function of the threshold $f$ and the system
size $N$.  (a)$\langle s^2 \rangle(f)$  for different system  sizes in
the  snooker model.  (b-d) Data  collapse obtained  by  expressing the
rescaled second moment of cluster sizes $N^{-\phi}\langle s^2 \rangle$
as function of the rescaled variable $(f-f_c)N^{1/2\nu}$ for the three
models  defined  in  the  text:  (b) snooker  model,  (c)  independent
$q$-model   and  (d)   microcanonic  $q$-model   The  values   of  the
corresponding  parameters $f_c$,  $\phi$ and  $\nu$ are  summarized in
Table  \ref{table}.  We do  not show  the data  for very  small system
sizes where the collapse takes place only in a small region around the
maximum.   For  the  $q$-models,  the  systems studied  had  the  same
vertical and horizontal linear sizes.  }\label{collapse}
\end{figure*}

The value of the critical threshold  $f_c$ depends on the scale set by
the external pressure. Under unit pressure, we found a different $f_c$
for each model.  Fig.   \ref{collapse} displays the scaling functions
obtained by  collapse of the data.  The estimated values of the
critical   thresholds   and  exponents   are   summarized  in   Table
\ref{table}, where the two-dimensional percolation exponents are also
included for reference.

In Fig.  \ref{phi}(b) we  show the probability distributions $P(F)$ of
force  magnitudes.  In  the independent  $q$-model, $P(F)$  is exactly
exponential \cite{fluct:jacco4},  while in the other two  models it is
exponential for large forces, and displays a peak at small forces.

\begin{table}
\caption{ Values of the critical threshold $f_c$ and the critical
  exponents $\phi$ and $\nu$ obtained from Fig.\ref{phi} and the data
  collapse shown in Fig.\ref{collapse}. For two-dimensional
  percolation, exact values are shown inside brackets.\label{table} }
\begin{tabular}{lccr}
 &$f_c$&$\phi=D-1$&$\nu$\\
\hline
Independent $q$-model &\hspace{1cm} $0.7\pm 0.01$ &\hspace{1cm} $0.69\pm 0.01$ &\hspace{1cm} $3.1\pm 0.1$\\
Snooker model &\hspace{1cm} $0.93\pm 0.01$ &\hspace{1cm} $0.89\pm0.01$ &\hspace{1cm} $1.65\pm 0.1$\\
Microcanonic $q$-model &\hspace{1cm} $0.585\pm 0.05$ &\hspace{1cm} $0.81\pm 0.01$ &\hspace{1cm} $1.65\pm 0.1$\\
Percolation &\hspace{1cm} &\hspace{1cm}$0.895 (43/48)$ &\hspace{1cm}$1.33 (4/3)$\\
\end{tabular}
\end{table}

\section{Discussion}
 We  have introduced  a new  approach to  investigate the  geometry of
force  networks, based  on statistics  of clusters  created  by forces
larger then a  given threshold. The existence of  a critical threshold
uncovers a scale-invariance of  force networks, which we characterized
by the critical exponents $\nu$  and $\phi$ for the correlation length
and  the   second  moment  of  the  cluster   size  distribution.   In
particular, in each network we  identify a fractal object of dimension
$D=\phi+1$,  given by  the  incipient force  cluster  at the  critical
threshold.  As  shown in Table  \ref{table}, we found  three different
sets of critical  exponents for the three models  we studied, implying
that  they  belong   to  distinct  geometrical  universality  classes,
although  their $P(F)$ display  similar features.   Interestingly, for
the snooker model,  $\phi$ is very  close to the  percolation value, but  as the
values  of $\nu$  are  further  apart and  the  scaling functions  are
different,  it does  not belong  to the  percolation universality
class.

Two distinct  universality classes could  have been expected  a priori
for  the  $q$-models  on one  hand  and  the  snooker packing  on  the
other.  Indeed, the  $q$-models  are both  directed  and include  only
scalar conservation  laws, while the  snooker model is  isotropic with
vectorial  conservation laws. The  reason for  the segregation  of the
independent and microcanonic in  two different universality classes is
more  subtle.  They  differ  only  by  the  form  of  the  probability
distribution of  forces, but in the independent  case the distribution
is Markovian from  top towards bottom, while in  the microcanonic case
no such preferred direction of propagation exists.

While in jammed matter the disorder in the underlying contact geometry
plays an important  role, we considered here only  lattice models with
fixed contact geometry.  The force-cluster  method can be applied in a
straightforward fashion to ensembles of forces networks resulting from
disordered contact networks. By analogy with other critical phenomena,
we however  do not expect  such randomness to modify  the universality
class. Indeed, the values we  have found for $\nu$ in combination with
the Harris' criterion \cite{harris} suggest that geometric disorder is
irrelevant.  A further study indeed shows that introducing quenched
disorder in the $q$-models does not modify the universality class,
which in turn confirms that the scale-invariance found here is in all
aspects similar to equilibrium critical phenomena.

While the results presented here clearly show that our method is able
to discriminate between different scaling behaviors, a crucial
question is whether any of the models belongs to the same universality
class as a realistic two-dimensional system of grains under isotropic
pressure. A recent study \cite{md} of packings generated by molecular dynamics
simulations showed that packings under isotropic pressure lead to the
same scaling behaviour irrespectively of the applied pressure, the
polydispersity of the grains, the coefficient of friction and the
force law. Remarkably, the corresponding scaling exponents and scaling
function appear to be the same as those obtained from the snooker packing.

\section{Outlook}

 The  existence of universality  classes for  force networks  raises a
number  of new  questions.  First  of all,  what properties  of a jammed
system determine  the universality class  of its force  network?  Our
results suggest  that that the isotropy  of the applied  force and the
vector nature of  the force balance are essential. On the other hand,
packings  under  static  shear  might  lead  to  another  universality
class. Another relevant parameter  could be the temperature in thermal
systems which exhibit jamming, such as colloids. 

Our  method based  on force  cluster  criticality is  clearly able  to
discriminate  between  the many  models  proposed  for force  networks
\cite{force:nicodemi,force:copper1,force:copper2,force:narayan,force:splitting,resp:gold3}.   In  particular,  it  shows  that  the
Edwards' hypothesis, which proposes  to consider all metastable states
of a  jammed system equally  likely, leads quantitatively to  the same
scaling properties as found in force networks generated by molecular
dynamics simulations.

Finally, the method developed here for force networks in jammed matter
is clearly  more general. It applies  in principle to  any ensemble of
graphs with continuous variables on the edges, such as flux, transport
or  metabolic networks  \cite{barabasi:fluxes1,barabasi:fluxes2}.  The
corresponding  universality classes  could complement  the topological
characterizations   of  networks   developed  in   the   recent  years
\cite{barabasi:rev}.

\bibliography{gran_stat}

\end{document}